\documentstyle[emulateapj]{article}

\tighten

\newcommand{\etal}{{\rm et al.~}}
\newcommand{\Mpc}{$h^{-1}$~{\rm Mpc}}
\newcommand{\hmpc}{$h$~{\rm Mpc$^{-1}$}}

\begin{document}

\title{Observational Matter Power Spectrum and the Height of the
Second Acoustic Peak}

\author{F. Atrio--Barandela} 
\affil{F{\'\i}sica Te\'orica. Facultad de Ciencias\\
        Universidad de Salamanca, 37008 Spain\\
        e--mail: atrio@orion.usal.es}

\author{J. Einasto}
\affil{Tartu Observatory, EE-61602 Estonia\\
        e--mail: einasto@aai.ee}

\author{V. M\"uller, J. P. M\"ucket}
\affil{Astrophysikalisches Institut Potsdam\\
        An der Sternwarte 16, D-14482 Potsdam, Germany\\
        e--mail: (vmueller, jpmuecket)@aip.de}

\author{A. A. Starobinsky}
\affil{Research Center for the Early Universe, University of Tokyo,
Tokyo 113-0033, Japan and \\
Landau Institute for Theoretical Physics, Moscow 117334, Russia\\
        e--mail: alstar@landau.ac.ru}

\begin{abstract}
We show that the amplitude of the second acoustic peak in the
newly released BOOMERANG-98 and MAXIMA-I data is compatible 
with the standard primordial
nucleosynthesis and with the locally broken-scale-invariant matter power
spectrum suggested by recent measurements of the power
spectrum in the range $20 - 200$~\Mpc.  If the slope of matter density
perturbations on large scales is $n \approx 1$, the Hubble constant is
$0.5< h < 0.75$, and r.m.s. mass fluctuations at 8~\Mpc\ are $0.65 \le
\sigma_8 \le 0.75$, then for a Universe approximately $14$~Gyr old our
best fit within the nucleosynthesis bound $\Omega_Bh^2= 0.019 \pm
0.0024$ requires $0.3 \le \Omega_m \le 0.5$. Cluster abundances further
constraint the  matter density to be $\Omega_m \approx 0.3$. The
CMB data alone are not able to determine the detailed form of the
matter power spectrum in the range $0.03<k<0.06$~\hmpc\ where
deviations from the scale-invariant spectrum are expected to be most
significant, but they do not contradict the existence of the
previously claimed peak at $k\sim 0.05$~\hmpc, and a depression at
$k\sim 0.035$~\hmpc.
\end{abstract}


\section{Introduction}

The recently released data from BOOMERANG-98 (de Bernardis et
al. 2000, hereafter B00) and MAXIMA-I (Hanany et al. 2000) have
determined the amplitude and position of the first and second acoustic
(Doppler) peaks of the Cosmic Microwave Background (CMB) radiation
angular spectrum. The first analysis (Balbi et al. 2000; Lange et
al. 2001; Jaffe et al. 2000) strongly constrained the parameter space
of cosmological models. The data clearly displayed a large first
acoustic peak with the maximum at $l=l_{max}=212\pm 7$ (Bond et
al. 2000) that supported the idea of an approximately spatially flat
Universe with a cosmological constant (vacuum energy) and non-baryonic
cold dark matter (CDM). However, rather unexpectedly, the second
acoustic peak at $l\sim 500 - 550$ appears to have a low amplitude
(especially in the B00 data). Another unexpected feature was a shift of
the first peak to smaller $\ell$ than the standard theoretical
prediction $l_{max}\simeq 220$ for the flat $\Lambda$CDM model. 
Even more recently, Netterfield et al. (2001, hereafter B01), 
Lee et al. (2001, hereafter M01) have 
estimated the radiation power spectrum up to $l\sim 1100$
by extending the analysis of BOOMERANG-98  and MAXIMA-I data
to smaller angular scales. Halverson et al. (2001, hereafter D01) presented
the angular power spectrum from the first season of DASI observations
with data out to $l\sim 800$.

A straightforward fit of the earlier data to $\Lambda$CDM models with a
scale-invariant initial spectrum of adiabatic perturbations lead to:

a) a high best fit value of the baryon density $\Omega_B h^2
\approx 0.03$ (see, e.g., Tegmark \& Zaldarriaga 2000b; Jaffe et
al. 2000) that significantly exceeds the prediction of the standard
Big-Bang nucleosynthesis (BBN), updated for the recent data on the
primordial deuterium abundance: $\Omega_B h^2 \approx 0.019\pm 0.0024$
(Tytler et al. 2000);

b) a best fit for the density $\Omega_{tot} \approx 1.1$ that
corresponds to a closed Universe (Lange et al. 2001, Jaffe et
al. 2000) (though the flat Universe is inside the $2\sigma$ error
bars).

The former result is mainly the consequence of the low amplitude of
the second acoustic peak, while the latter is mostly due to the shift
of the first peak to the left. These conclusions persist if additional
information on the Large Scale Structure (LSS) of the Universe, as the
density fluctuation parameter $\sigma_8$, and the shape of the power
spectrum of matter are taken into account. In particular, the most
exhaustive of these efforts made by Tegmark, Zaldarriaga \& Hamilton
(2000), who performed an 11-parameter fit to the current CMB and LSS
data, has led to essentially the same result about the high baryon
density.
To avoid contradiction with the standard BBN, a number of drastic
changes in the standard FRW cosmology were proposed, as leptonic
asymmetry (Lesgourgues \& Peloso 2000), delayed recombination
(Peebles, Seager \& Hu 2000), loss of coherence of primordial
perturbations remained from inflation (White, Scott \& Pierpaoli
2000), admixture of topological defects (Bouchet et al. 2000) or even
the absence of dark non-baryonic matter (McGaugh 2000).  Moreover,
even the fundamental laws of physics itself, as the constancy of the
fine structure constant, have been abandoned for the sake of
explaining these features (Avelino et al. 2000, Battye et
al. 2001).  

A preliminar study of the newest D01 data shows no
contradiction with the BBN bound; B01 agrees with the earlier
analysis of Lange et al (2001) except that the baryon abundance is
reduced to $\Omega_bh^2 = 0.027\pm 0.005$. This limit gets
even closer to the BBN estimates if all points are considered in the
analysis. But at high multipoles, the data is not fully
consistent. M01 data show a third peak higher than the second, while in
B01 and D01 both acoustic peaks are of similar height. 

All studies quoted above, even those including information on LSS,
considered matter power spectra only  with scale-invariant initial conditions.
But Einasto \etal (1999c) have shown that the matter power
spectrum as measured from galaxy and cluster catalogs is inconsistent
with this assumption. A number of different non-scale-invariant
initial conditions has been recently used to analyze the CMB data.
First, Kanazawa et al. (2000) considered a double inflationary model
in supergravity having a step in the spectrum located around
$k=0.03$~\hmpc, for which both the matter power at smaller scales and
the amplitude of the second Doppler peak are reduced. A similar matter
spectrum was  studied in Barriga et al. (2000) which resulted from
another kind of inflationary model with a fast phase transition during
inflation. In contrast, Griffiths, Silk \& Zaroubi (2000) and, most
recently, Hannestad, Hansen \& Villante (2000) advocated the existence
of a bump in the matter spectrum on significantly larger scales
($k\approx 0.004$~\hmpc) based on purely phenomenological grounds.
Einasto (2000) analyzed the CMB spectrum resulting from a matter power
spectrum with a bump at $k= 0.05$~\hmpc\ as suggested by Chung et
al. (2000).

In most of these papers, the location of non-scale-invariant features
in the spectrum, or even the functional form of these features were
introduced {\it ad hoc}, with the only aim to explain the
BOOMERANG-MAXIMA early data. In contrast, we adopt a completely different
approach: without attaching ourselves to any particular theoretical
model, we are trying to use previously existing observational data as
much as possible. To obtain the initial (post-inflation) matter
spectrum, first, we use the empirical present matter power spectrum of
Einasto \etal (1999a), extracted from galaxy and cluster catalogs and
estimated in the range $\sim 20 - 200$~\Mpc. Secondly, on very large
scales we assume the initial spectrum to be scale-invariant ($n=1$)
and COBE/DMR normalized, since this choice produces the best fit to
previous CMB data for multipoles $l<200$ (fortunately, it agrees with
the prediction of the simplest version of the inflationary scenario,
too). In the next section we shall describe two possible ways to match
these pieces of the spectrum.

In this article, we show that such an empirical spectrum gives a
possibility to explain the peculiar features of the BOOMERANG-98 and MAXIMA-I
data without changing standard cosmology 
or the basic
laws of physics.  Also, we do not introduce non-scale invariant
features in the power spectrum {\it ad hoc}, but only those required
by the observations.  In addition to this initial spectrum, we assume
certain priors: a spatially flat universe, the age of the universe
between 12 and 14 Gyr and a negligible contribution from primordial
gravitational waves to the COBE/DMR data.  We considered models with
different Hubble constant, cosmological constant, and baryon fraction.
We computed the expected temperature anisotropy for each model and 
found the region of the parameter space that best fitted the
BOOMERANG-MAXIMA data. In Section 2 we describe our methods and
results. In Section 3 we find the cluster mass distribution. We
calculate the initial power spectrum for our models in Section 4 and
present the main results in Section 5.

\section{Temperature Anisotropies on Intermediate Angular Scales}

Temperature anisotropies are usually expressed in terms of spherical
harmonics with coefficients $a_{lm}$.  At a given angular scale $l$,
the contribution of the matter power spectrum to the radiation
temperature anisotropy is dominated by perturbations with the wavenumber
$k\simeq (\pi/2) l/R_H$, where $R_H$ is the radius of the present
(post-inflation) particle horizon (in particular, $R_H=9900$~\Mpc\ for
a flat Universe with the matter density parameter $\Omega_m=0.3$). Note
that only perturbations with wavenumbers $k>l/R_H$ contribute to the
CMB radiation temperature at any $l$ (for $l\gg 1$ and neglecting the
integrated Sachs-Wolfe effect which is usually small for scales of
interest here).  Observations in the range $l\ge 200$ probe the
amplitude of the matter power spectrum at recombination on the scales
also probed by galaxy and cluster catalogs, i.e. on scales smaller
than 200~\Mpc.  On very large scales ($k\le 0.005$~\hmpc), temperature
anisotropies are generated by fluctuations of the gravitational
potential. The amplitude and the slope of the matter power spectrum in
this region are strongly constrained by the COBE/DMR data.  To compute
CMB temperature anisotropies for $l = 2-1000$, and the matter power
spectrum on scales $10^{-4} \le k \le 1$~\hmpc, we need to provide the
initial (post-inflation) power spectrum in the whole wavenumber range.

We identify the matter power spectrum with the observational spectrum
presented in Einasto et al. (1999a) (called HD for High Density as the
mean power spectrum was derived including high-density regions such as
clusters of galaxies) for $k\ge 0.035$~\hmpc.  This spectrum is well
approximated by a power law with the slope $n=-1.9$ for wavenumbers
larger than $k_{pl} = 0.06$~\hmpc\ up to $k\sim 0.4$~\hmpc (the latter
value is already sufficient for the calculation of CMB multipoles with
$l<1000$, so we use this fit for larger values of $k$, too):
\begin{equation}
P_{HD}(k) = P_{pl}~(k/k_{pl})^{-1.9}~h^{-3}{\rm Mpc}^3,
\label{spec}
\end{equation}
where $P_{pl}=1.02\times 10^4 ~(\sigma_8/0.65)^2 ~h^{-3}{\rm Mpc}^3$
is the value of the power spectrum at $k=k_{pl}$ (see Table 2, Einasto
et al. 1999a).  In the range $0.035 \le k \le 0.3$~\hmpc\ this
spectrum differs from the observed power spectrum of galaxies, by a
scale factor or bias.  On shorter scales, $k\ge 0.3$~\hmpc, the
observed spectrum is corrected for non-linear evolution effects (for
details see Einasto et al. 1999c); these corrections do not introduce
any significant differences to the CMB spectra at the scales of
interest.  Note that in this range, the spectrum of Einasto et
al. (1999a, 1999c) is in good agreement with results obtained by other
authors, for recent new data see Miller et al. (2001), among
others. Since we do not know the exact value of the bias between the
galaxy and matter power spectra, the value of $k_{pl}$ is a free
parameter of our model which is expressed in terms of $\sigma_8$ once
the shape of $P(k)$ is fixed.  On the other hand, on very large
scales, $k<k_m=0.03$~\hmpc, we accept the scale-invariant ($n=1$) COBE
normalized spectrum since it gives the best fit to CMB data for
$l<200$.  The location of $k_m$ is chosen to be in agreement with the
observational cluster spectrum of Miller \& Batuski (2000) and the
de-correlated IRAS Point Source redshift catalog (PSCz) galaxy
spectrum of Hamilton, Tegmark \& Padmanabhan (2000).

\begin{figure*}[t]
\vspace*{15cm}
\includegraphics{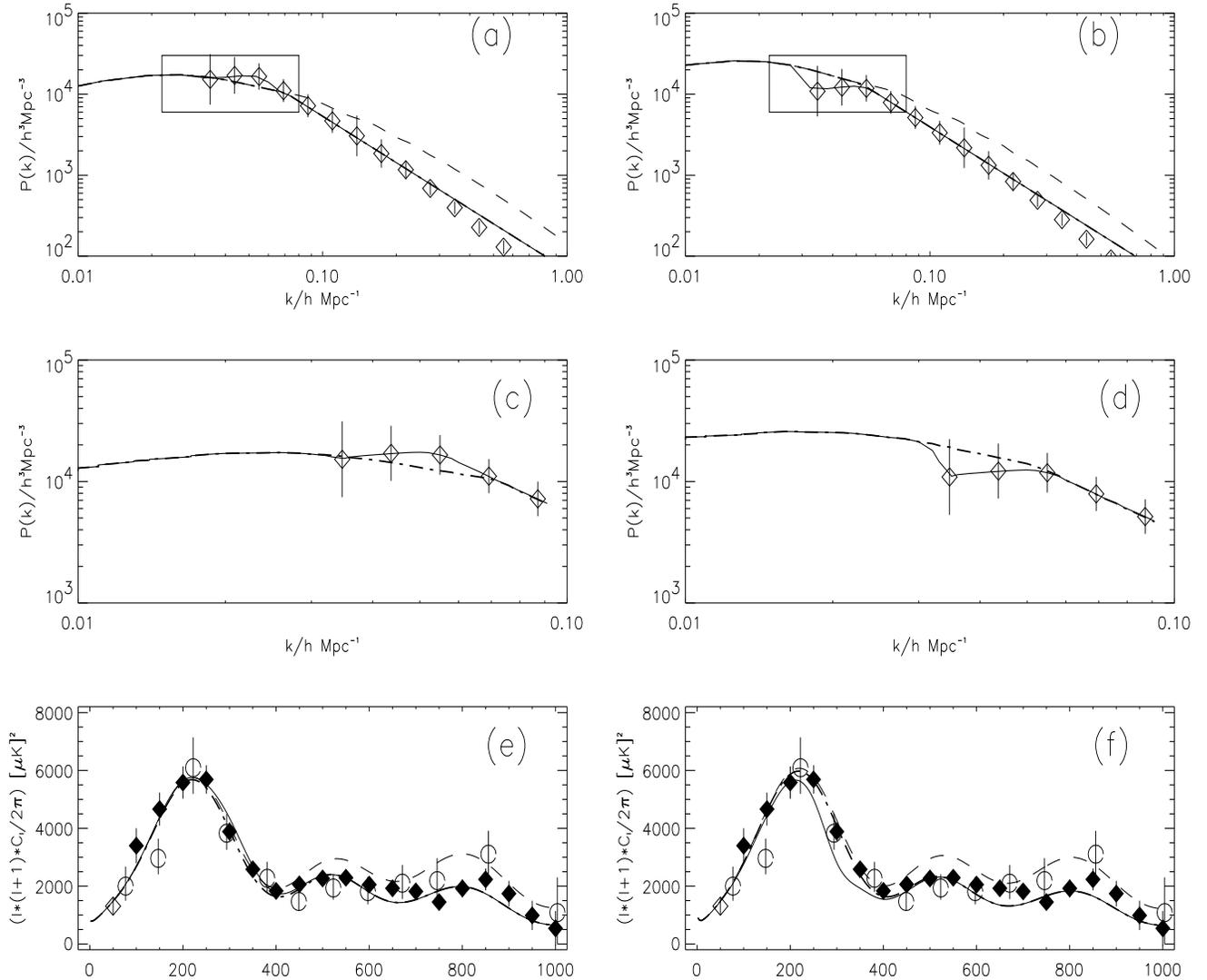}
\caption{(a,b) Matter power spectrum of $\Lambda$CDM models with scale
invariant (dashed line), HDB (solid line) and HDL (thick dot-dashed line)
initial conditions. The matter density is $\Omega_m = 0.5$ in (a) and
$0.3$ in (b).  The box encloses the region where the observed power
spectrum is extended to match smoothly the scale invariant spectrum at
large scales.  Diamonds and bars represent the observed matter power
spectrum $P_{HD}(k)$ and the associated $1\sigma$ error bar (data
taken from Table 2 of Einasto et al 1999a). This area has been
enlarged in (c,d) to show the difference between the spectra HDL and
HDB. For clarity the scale invariant $\Lambda$CDM model is not plotted.
(e,f) Radiation power spectrum of the same models. Filled
diamonds and open circles correspond to B01 and M01 data, 
respectively.}
\end{figure*}

In the intermediate range, $k_m \le k \le k_{pl}$, error bars are
large and there is no complete agreement between different authors
about the exact form of $P(k)$. For this reason, we use two different
models of $P(k)$ in this region: one more conservative and another
based on the Einasto et al. (1999a) data.  In the first case we use at
small scales the power law behavior given in (\ref{spec}).  This
spectrum is extrapolated until it crosses the linear COBE-normalized
spectrum that is based on the primordial scale invariant spectral
index $n=1$ and evolved through recombination using cosmological
parameters discussed above.  On large scales the spectrum based on is
$n=1$ used. It was calculated using the CMBFAST program. This power
spectrum is called HDL (L standing for linear extrapolation). In the
second case we apply the observational HD spectrum from Einasto et
al. (1999a) up to $k=0.035$~\hmpc, i.e., up to the smallest wavenumber
for which it contains reliable data.  Then $P(k)$ is linearly matched
to the scale invariant $n=1$ COBE normalized spectrum at
$k=k_m=0.03$~\hmpc. For smaller $k$, we use the scale invariant
spectrum as in the first case. The resulting spectrum has a distinct
bump at $k=k_b$, and a depression in the whole region $k>k_m$, as
compared to the COBE normalized $n=1$ spectrum, see Fig~1a,b. Thus we
call it HDB (B for bump). Notice that the difference between the two
spectra is still within $1\sigma$ error bars.

\begin{figure*}[t]
\vspace*{12cm}
\includegraphics{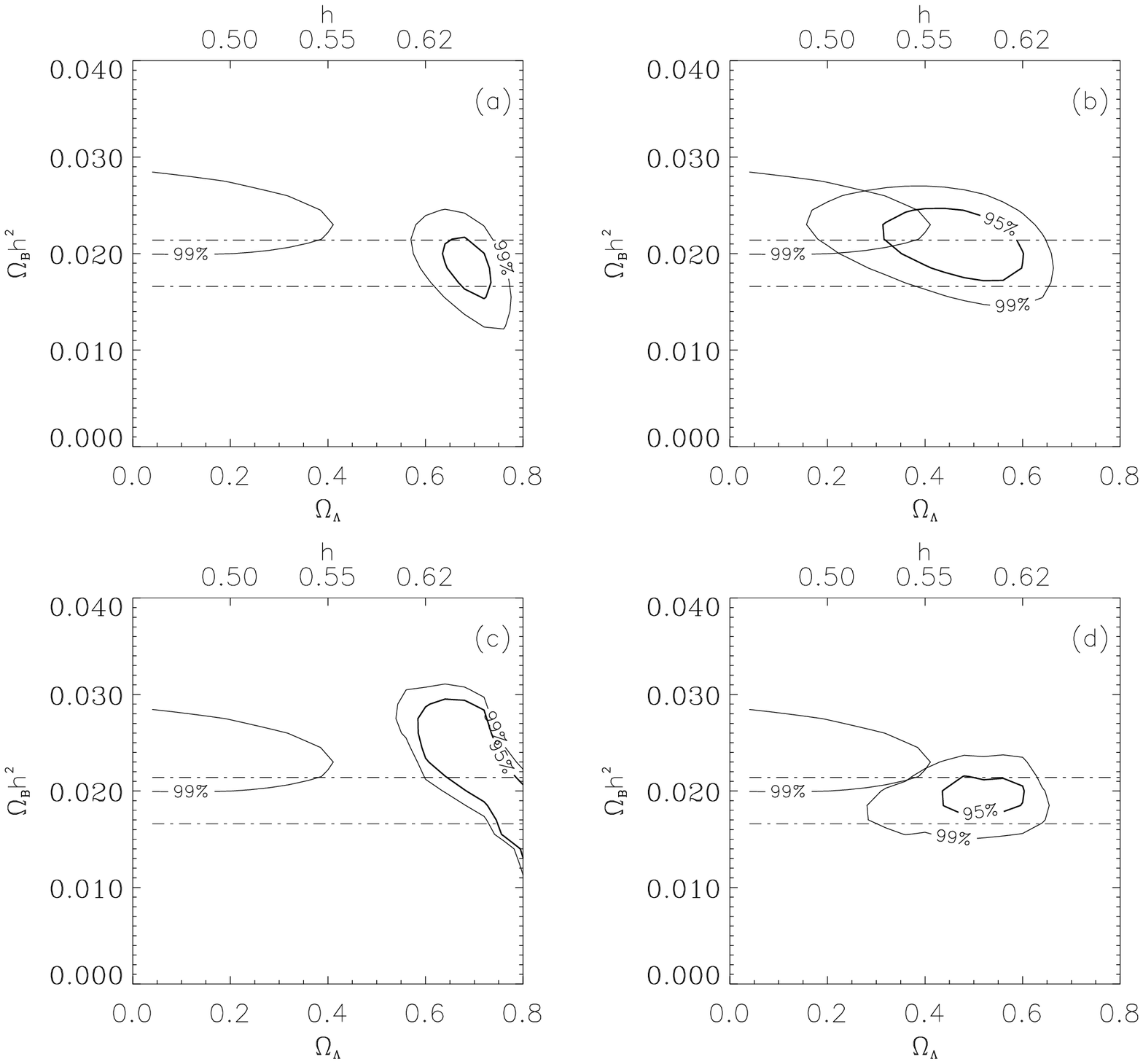}
\caption{Goodness-of-fit contours of the likelihood function at 95\%
(thick solid line) and 99\% (thin solid lines) confidence
levels.  The fit was performed using
the newly released data B01, M01 up to $l\simeq 650$. No calibration
uncertainties were included.
Contours on the left side of  panels correspond to scale invariant
$\Lambda$CDM models and are shown for comparison; right contours
correspond to HDL or HDB models.  All models
correspond to a Universe age of 14 Gyr and to matter power spectra
with a $n=1$ spectral index on large scales.  Upper two panels: solid
lines correspond to a HDL power spectrum with (a) $\sigma_8=0.65$ and
(b) $\sigma_8=0.75$. Lower two panels: solid lines correspond to a HDB
power spectrum with (c) $\sigma_8=0.65$ and (d) $\sigma_8=0.75$.
Dot-dashed straight lines lines correspond to the 1$\sigma$ BBN limits
on the baryon fraction. 
}
\end{figure*}

The HDL spectrum is similar to those considered in Kanazawa et al.
(2000) and Barriga et al. (2000). The HDB spectrum is more
complicated, it is not monotonous.  Its form is in agreement with the
power spectrum proposed by Einasto et al. (1997); the existence of the
bump at $k=0.05$~\hmpc\ (now understood to be a relative bump above a
depression in the spectrum) could help to explain some quasi-periodic
features in the large-scale distribution of Abell clusters noted in
the latter and other papers. It is interesting that the double feature
-- the bump at $k=0.05$~\hmpc\ and the depression at $k\simeq
0.035$~\hmpc\ -- are seen in the Miller \& Batuski (2000) and
Hamilton, Tegmark \& Padmanabhan (2000) data, too, though within
$1\sigma$ error bars.  Similar features have been reported in two
recent preprints: Hoyle et al. (2001) found a small bump at $k\approx
0.06$~\hmpc\ (assuming $\Omega_m=0.3$) in the power spectrum of the
2dF QSO Redshift Survey, and Silberman et al. (2001) claimed a wiggle
in the power spectrum with an excess at $k\sim 0.05$~\hmpc\ and a
deficiency at $k\sim 0.1$~\hmpc, based on peculiar velocities of
galaxies from the Mark III and SFI catalogs. Bumps at $k\approx
0.05$~\hmpc\ and $k\sim 0.2$~\hmpc\ and depressions (valleys) at
$k\simeq 0.035$~\hmpc\ and $k\sim 0.1$~\hmpc\ have been confirmed by
Miller, Nichol \& Batuski (2001) in their analysis of power spectra of
Abell and APM clusters of galaxies, and PSCz galaxies.  Note, however,
that there are no such features in the recent REFLEX X-ray cluster
survey (Schuecker et al. 2000).  Certainly, both HDL and HDB are
non-scale-invariant spectra. Though we have arrived at them using
purely empirical arguments, we shall discuss how they could have
arisen in inflationary models in Section 4.

To compute the radiation and matter power spectrum at the present
epoch we used the CMBFAST program developed by Seljak \& Zaldarriaga
(1996). Given a set of parameters and scale-invariant initial
conditions of the type $\tilde\delta(k) = Ak^{(n-1)/2}$ for the rms
Fourier amplitudes of the density contrast at horizon crossing, we
obtained a matter power spectrum with spectral index $n$ at large
scales. Parameter $A$ was normalized to reproduce the COBE/DMR
amplitude.  In this work, we assumed that a difference between the
power spectrum predicted by a given cosmological model today, $P(k)$,
and the observed power spectrum, $P_{HD}(k)$, was due to initial
conditions only. Then, as all scales of interest are still in the
linear regime, the initial rms Fourier amplitude of matter density
perturbations that give rise to the observed spectrum is
\begin{equation}
\tilde\delta_{HD}(k) = \tilde\delta(k)(P_{HD}(k)/P(k))^{1/2}~.
\label{ini}
\end{equation}
In this expression, the power spectra are evaluated at the present
time, and the amplitude of $\tilde\delta_{HD}(k)$ is evaluated at the
horizon crossing.

\begin{center}
\begin{table*}[t]
\caption{{Position of the first maximum of the CMB spectrum}}
\label{tab:param1}
\begin{tabular}{lcc}
\\ \hline \hline \\
%
Model          & $\Omega_m = 0.5$  &  $\Omega_m = 0.3$\\
 \\ \hline \\
Scale invariant&         219    &         217  \\
HDL            &         218    &         216  \\
HDB            &         214    &         209  \\
\ \\ \hline   
\end{tabular} 
\end{table*}
\end{center}

Our main goal is to determine if the B01 and M01 new measurements agree
with the observed matter power spectrum and with the standard BBN.
Therefore, we restricted our analysis of the parameter space to a
rather limited set, centered upon the region that best fitted the CMB
data before BOOMERANG-98 and MAXIMA-I data
(see Tegmark \& Zaldarriaga 2000a).  We
considered spatially flat models with and without cosmological
constant ($\Omega_m+\Omega_\Lambda = 1$, $\Omega_m = \Omega_{CDM}+
\Omega_B$), with matter density in the range $0.3\le\Omega_m \le 1$,
with $n=1$ and baryon fraction $0.005\le\Omega_Bh^2\le 0.035$ centered
on the range suggested by Tytler et al. (2000).  Particular attention
was paid to models with a cosmological constant. We took the age of
the Universe $t_0$ to be between $12$ and $14$~Gyr, in agreement with
the recalibration of the cosmic ages made by Feast \& Catchpole (1997)
using new Hipparcos distance determinations, and from estimations of
the age of globular clusters (Jimenez 1999). The resulting Hubble
constant varied from $h\simeq 0.5$ to $0.75$, consistent with
observations.

Models with a small mixture of massive neutrinos have also been
discussed in the literature, but this component has little effect on
the radiation power spectrum in the range considered here. However, it
is important to realize that as neutrino free-streaming erases the
matter power spectrum at small scales, they have an effect on the
initial matter power spectrum computed using (\ref{ini}).  If the dark
matter is ``cold'' and $\Omega_m \simeq 1$ then the initial spectrum
deviates strongly from scale invariant initial conditions.  This is
expected since the standard CDM has too much power on small scales.
If we assume that the observed spectrum deviates only weakly from the
Harrison-Zeldovich spectrum, then the theoretical power spectrum on
scales smaller than the bump ($k \ge 0.05$~\hmpc) coincides with the
observed spectrum only for a restricted set of cosmological
parameters.  In particular, demanding that the observed spectrum
deviates only slightly from scale invariance, we fix the matter
density parameter to be $\Omega_m \approx 0.3$ (if the non-baryonic
dark matter is ``cold''). A small mixture of ``hot'' dark matter would
allow for a larger matter density (Einasto 2000 and Section 4 below).

In Figure~1, we present the COBE-normalized matter power spectra
(upper panels a and b with an enlarged version in the middle panels c
and d) and radiation power spectra (lower panels e and f) for three
different types of spectra: scale invariant (dashed line), HDB
(solid) and HDL (thick dot-dashed), for a universe with $t_0 = 14$~Gyr, the
baryon fraction $\Omega_Bh^2 = 0.02$ and the matter density $\Omega_m
= 0.5$ (left panels) and $\Omega_m = 0.3$ (right panels). In this
figure we use $\sigma_8 = 0.75$ in the left panel and $\sigma_8 =
0.65$ in the right panel.  
The HDL power
spectrum follows Eq~(\ref{spec}) till $k\simeq 0.08$~\hmpc (for
$\Omega_m=0.5$), and till $k\simeq 0.06$ (for $\Omega_m=0.3$) where it
follows the $\Lambda$CDM spectrum. Then HDL coincides or is below the
scale invariant power spectrum. The HDB spectrum follows the observed
spectrum up to $k=0.035$~\hmpc, and then is linearly extrapolated
until it reaches the scale invariant spectrum at $k = 0.03$~\hmpc. It
can be above or below the scale invariant spectrum.  
In the radiation spectra, the main differences between HDB 
and HDL are around the first Doppler peak. The data on the matter
power spectrum was taken from Einasto et al. (1999a), while
the CMB data is the latest M01 and B01 (this is also the data
used to fit the models shown in Figure~2).

\begin{figure*}[t]
\vspace*{8cm}
\includegraphics{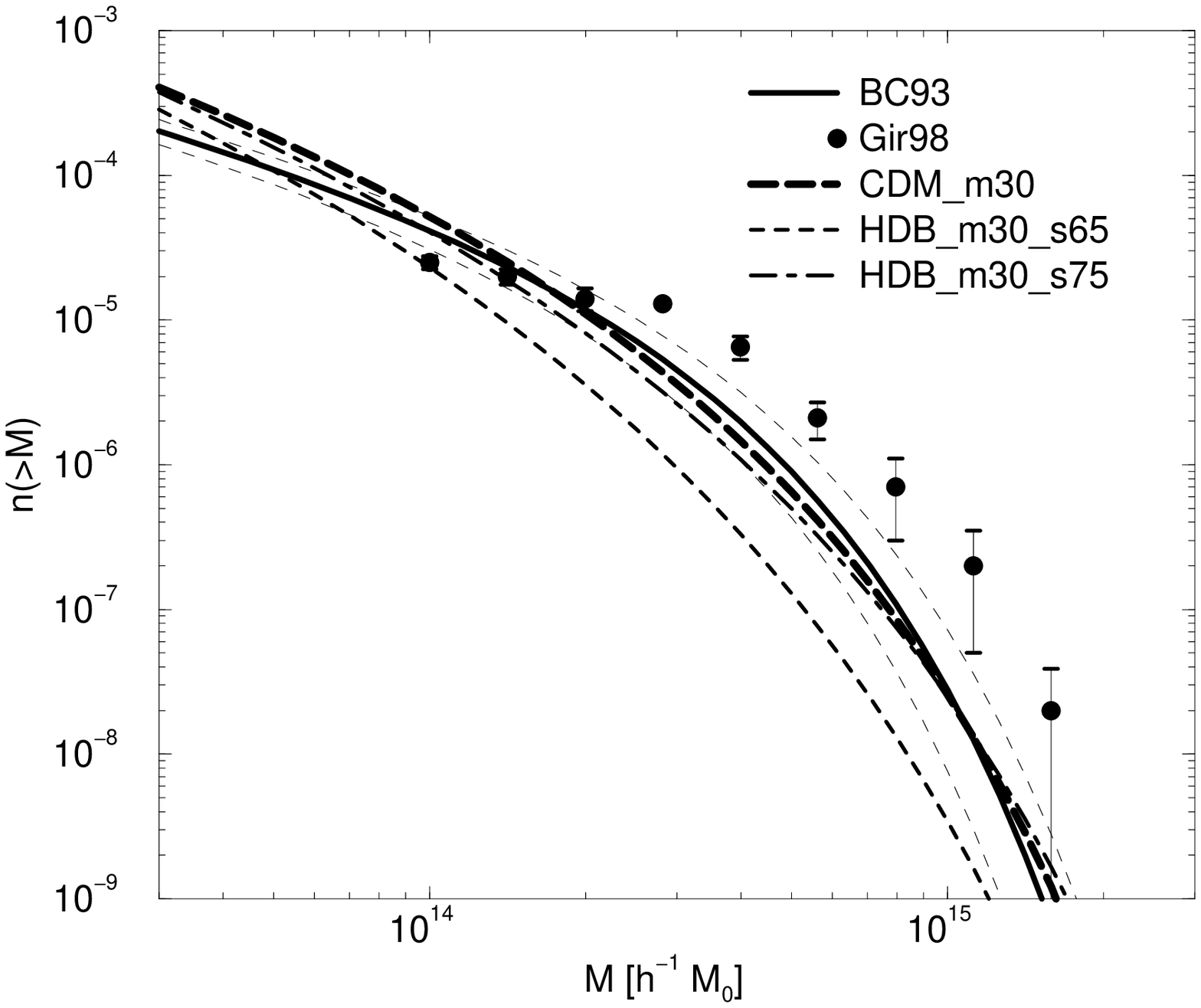}
\includegraphics{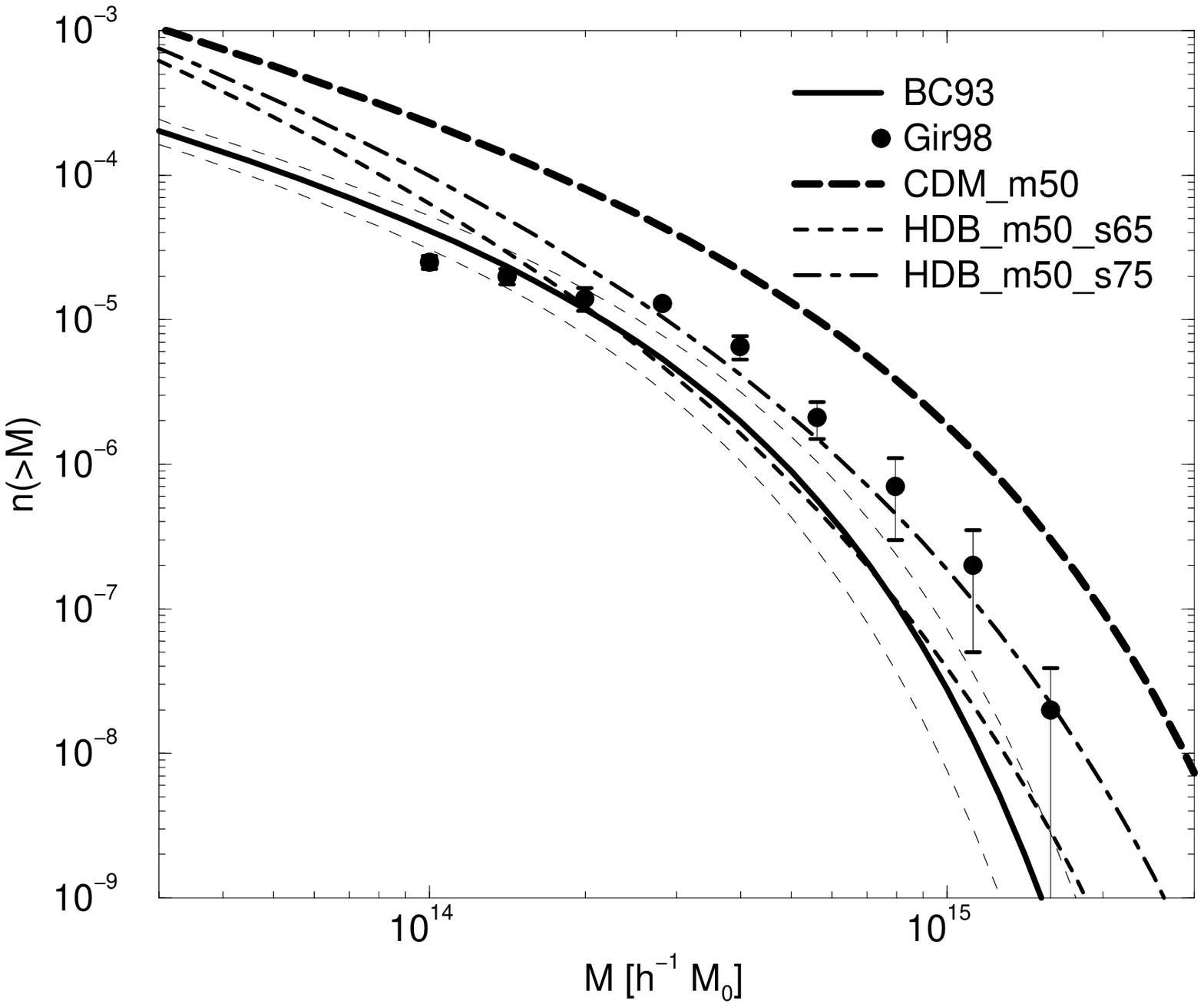}
\caption{Cluster mass functions for models with the density parameter
$\Omega_m=0.3$ and $\Omega_m=0.5$ are plotted in the left and right
panels, respectively. The observed cluster mass functions are given
according to Bahcall \& Cen (1993) (thick solid line indicates the
data and thin solid lines the width of the error corridor) and Girardi
\etal (1998).  Model functions are given for scale-invariant CDM
models (with a cosmological constant), and for our models HDB with a
bump for $\sigma_8=0.65$ and $\sigma_8=0.75$.}
\end{figure*}

From the initial power spectrum, we compute the radiation power
spectrum and compare the expected level of anisotropy with the CMB
data. We used the radical compression of the cosmic microwave
background data method described in Bond, Jaffe \& Knox (2000). We computed
the parameter space compatible with the newly 
released B01, M01 and,  for comparison, with 
the earlier data recalibrated according to
Jaffe et al. (2000). We restricted our study to $l\le 650$ 
since  it allows a better comparison with earlier parameter
estimates. Further, above that scale the error bars are most affected by
$\pm 13\%$ effective beam uncertainty
and this uncertainty is not included in the published error bars.
In Fig~2 we give the contours at the 95\%
($2\sigma$) and 99\% confidence levels for $\Lambda$CDM models with
scale invariant, HDB and HDL power spectra. The figure corresponds
to a fit to the B01, M01 data.  On the upper axis we plot
the corresponding Hubble constant. All models presented in Fig.~2
correspond to the slope $n=1$ on very large scales.  Figure~2a shows
that scale invariant models are marginally consistent with the BBN
bounds only at the 99\% level.  Our results also indicate that HDL and
HDB models with $t_0=12$~Gyr (not shown in the figure) require
$\Omega_{\Lambda}>0.75$ (at the $2\sigma$ level), hardly compatible
with direct measurements of $\Omega_m = 1- \Omega_{\Lambda}$.  On the
contrary, models with $t_0=14$~Gyr lay inside the standard BBN limits
and fit the recent CMB data very well for both the HDL and HDB
spectra. We repeated the same analysis with the original recalibrated
data. In this case, the contours were wider and scale invariant
models become compatible with the BBN bound.
Let us remark that, as seen in Figure~1, BOOMERANG-98 
and MAXIMA-I have different amplitudes at the scale of the second
acoustic peak, thus it is harder for the models to fit them
both. However, in both cases the contours for HDL and HDB
spectra were centered on the same region of the
parameter space, implying that our results are robust.

In Table~1 we give the position of the maximum of the first acoustic
peak for the models presented in Fig.~1.  Let us remark that since, in
general, the HDB and HDL spectra have less power than a scale
invariant model with the same cosmological parameters, the position of
the maximum of the first acoustic peak is shifted with respect to the
scale invariant model.  This effect is easily explained by the $k-l$
correspondence given at the beginning of this section: the depression
in the HDB spectrum relative to the scale-invariant spectrum in the
range $k=0.03-0.05$~\hmpc\ results in the decrease of the
corresponding angular radiation spectrum for $l=200-500$ (if
$\Omega_m=0.3$). This effect explains the position of the first
Doppler peak without invoking a small positive spatial curvature.

\section{Cluster mass function}

Another important constraint of cosmological models is the cluster
mass function.  This function is rather sensitive to cosmological
parameters and features in the power spectrum on intermediate and
small scales which have the highest weight in the cluster formation
process.  To check our results we calculated the cluster mass function
using the Press-Schechter (1974) method.  Mass functions were found
for two sets of models; we used the Hubble constant $h=0.65$ and the
baryon fraction $\Omega_B=0.05$, the matter density was taken
$\Omega_m=0.3$ and $\Omega_m=0.5$, varying the density of the cold
dark matter and the vacuum energy (cosmological constant)
respectively.  The Hubble constant used is a compromise between recent
new estimates by different teams (Parodi et al. 2000, Sakai et
al. 2000).  As before, we used only spatially flat models ($\Omega_m +
\Omega_{\Lambda} =1$); the slope of the spectrum on small scales was
taken to be $n=-1.9$, and the spectrum amplitude parameters
$\sigma_8=0.65, ~0.75$ were used.

The results of calculations are shown in Figure~3.  The HDB model with
$\Omega_m=0.3$ and $\sigma_8=0.65$, that fits the CMB data rather
well, has a too low abundance of clusters while the scale invariant
$\Lambda$CDM model and HDB model with $\Omega_m=0.3$ and $\sigma_8=0.75$
lie within the range allowed by observations.  It is not surprising
that the last two models coincide since the $\Lambda$CDM model with
$\Omega_m=0.3$ has $\sigma_8=0.78$, very close to our HDB model with
$\Omega_m=0.3$ and $\sigma_8=0.75$.  This $\sigma_8$ value is in good
agreement with other independent estimates (Einasto \etal 1999b, Van
Waerbeke \etal 2001).  The $\Lambda$CDM model with $\Omega_m=0.5$ has
too high abundance of clusters over the whole mass range (about a
factor of ten).  In contrast, $\Omega_m=0.5$ HDB models with both
values of $\sigma_8$ have cluster mass functions which lie within the
allowed region.  In other words, cluster mass functions of HDB models
are rather insensitive to the density parameter.  This result is
expected since our HDB models with both density values (but an
identical Hubble parameter) are identical on medium and small scales,
$k \ge 0.06$, which have the highest weight in determining the cluster
mass function.

\begin{figure*}[t]
\vspace*{8cm}
\includegraphics{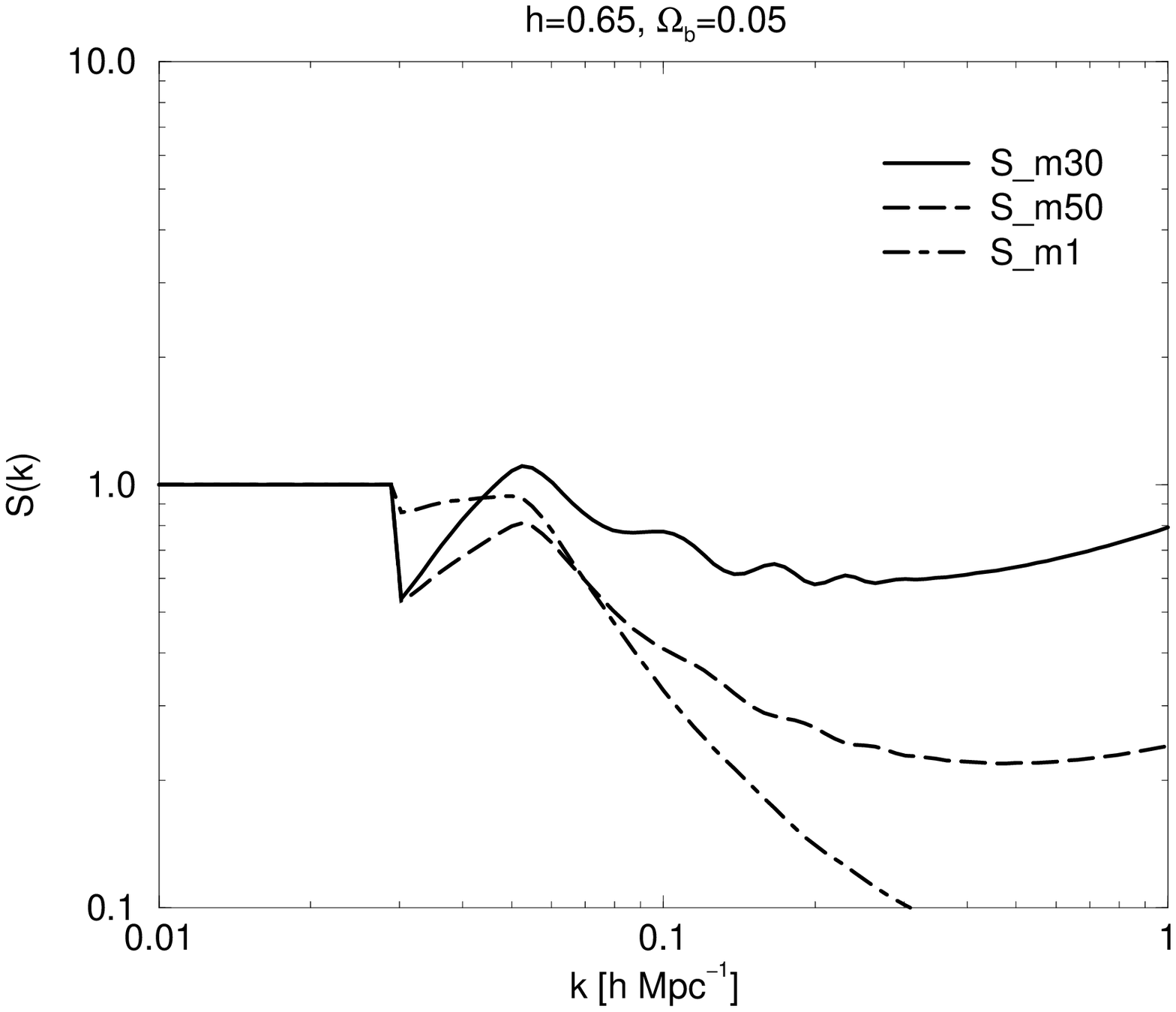}
\includegraphics{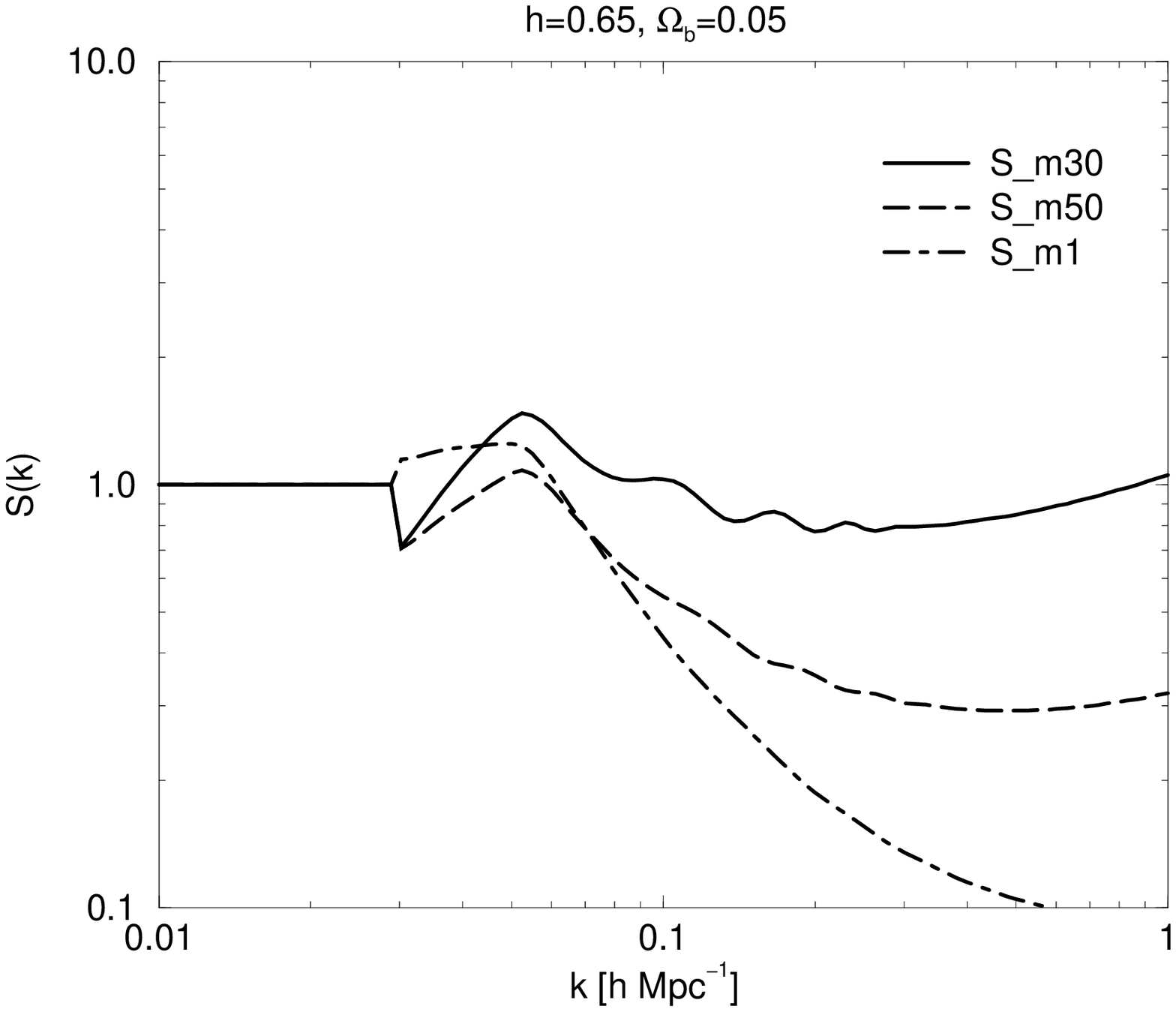}
\caption{The initial power spectrum, expressed as the ratio of our adopted
HDB spectrum to the scale-invariant $\Lambda$CDM  spectrum,
$S(k)=P_{HDB}(k)/P_{CDM}(k)$.  The initial power spectrum is calculated
for 3 values of the matter density, $\Omega_m =0.3,~~0.5,~~1.0$. In
the left panel  $\sigma_8=0.65$, in the right
panel $\sigma_8=0.75$.}
\end{figure*}

Using the preprint version of this paper, Gramann \& H\"utsi (2001)
investigated properties of models similar to our models HDL, fixing
the break of the power spectrum at $k_{pl}=0.05$~\hmpc\ (instead of at
$0.06$~\hmpc).  They confirmed our results that the amplitude of the
second and third peak in the radiation power spectrum decreases and
fits the CMB data better.  Gramann \& H\"utsi also calculated the
cluster mass function with the Press-Schechter method and found that
the amplitude of the power spectrum of their models was too low which
lead to a low cluster mass function incompatible with observations.
They suggested to use an initial power spectrum which has a depression
around $k=0.1$~\hmpc\ and an increased amplitude starting from
$k=0.2$~\hmpc.  Models with such spectra were investigated by
Suhhonenko \& Gramann (1999).  Such spectra also yield satisfactory
agreement both with the observed cluster mass function and with the
recent CMB spectrum data.  A similar power spectrum was found in the
recent paper by Silberman et al.  (2001) based on the analysis of
galaxy peculiar velocities, and by Miller, Nichol \& Batuski (2001)
based on the analysis of power spectra of Abell and APM clusters of
galaxies, and IRAS Point Source redshift catalog galaxies.

To bring our HDB model with $\Omega_m=0.3$ and $\sigma_8=0.75$ into a
better agreement with the CMB data, perhaps a model with a second
feature on small scales is needed, as suggested by Gramann \& H\"utsi
(2001) and Miller, Nichol \& Batuski (2001). The latter authors
considered the presence of two bumps and two depressions (valleys) in
the observed power spectrum as evidence for baryonic fluctuations, as
discussed by Eisenstein \etal (1998).  However, this hypothesis
encounters one problem: the position of the first baryonic bump on the
matter power spectrum is expected at the wavenumber $k \approx
0.07$~\hmpc, whereas the observed bump is located at the wavenumber $k
\approx 0.05$~\hmpc.  The bump at $k =0.05$~\hmpc\ corresponds to the
scale of the supercluster-void network, $\sim 120$~\Mpc, as seen in
the distribution of high-density regions in a pencil-beam near
galactic poles (Broadhurst et al. 1990), and in the distribution of
clusters of galaxies (Einasto et al. 1997a, 1997b).  More accurate
data are needed to decide whether the observed features of the power
spectrum are due to a high baryon fraction or are primordial effects.
At present, we are inclined to accept the second alternative since,
without requiring new physics, it fits the CMB data within the BBN
bounds quite naturally.

\section{Post-Inflation Initial Conditions}

Though we did not require {\it a priori} that our initial spectra
should be ``explicable'' by any theoretical model, it appears that,
for $\Omega_m \approx 0.3$, the form of the HDB spectrum suggests an
inflationary model with a fast phase transition occurring
approximately $50$ e-folds before the end of inflation, similar to
those with a sudden jump in an inflaton potential which were
considered in Starobinsky (1992) and Adams, Ross \& Sarkar (1997); see
also the recent paper by Adams, Cresswell \& Easther (2001) (which
appeared after the preprint version of the present paper was sent to
the archive) where a very similar spectrum is obtained. As for the
simpler HDL spectrum, and for the HDB spectrum with $\Omega_m
\approx 0.5$, it is rather typical for double inflationary
models, see, e.g., the models in Polarski \& Starobinsky (1992),
Gottl\"ober, M\"ucket \& Starobinsky (1994), as well as the models
proposed in Kanazawa et al. (2000) and Barriga et al. (2000). Also,
a similar step-like spectrum appears in the case when the inflaton
potential has a jump of its first derivative (Starobinsky 1992).
Thus, relatively small changes in cosmological parameters and in the
present form of $P(k)$ may lead to significantly different models
of the phase transition during inflation.

To elaborate this point further, let us compute the ratio of power
spectra: 
\begin{equation}
S(k) = P_{HDB}(k)/P_{CDM}(k)~.
\label{init}
\end{equation}
Here, we use the power spectrum with a bump, $P_{HDB}(k)$, and the
scale-invariant $\Lambda$CDM power spectrum $P_{CDM}(k)$, calculated
for the same set of cosmological parameters as the spectrum with a
bump.  The function $S(k)$ characterizes the deviation of our accepted
spectrum from the scale-invariant case, namely, the initial spectrum
$P_0(k)\propto kS(k)$. In Fig.~4 we plot the ratio $S(k)$ for a Hubble
constant $h=0.65$, baryon density $\Omega_B=0.05$, and $\sigma_8=0.65$
and $\sigma_8=0.75$ for three different values of the matter
density. We do not plot the ratio using the HDL spectrum since $S(k)$
is the same as for HDB for $k\ge 0.06$~\hmpc\, and coincides with
the scale invariant case ($S(k)=1$) for $k \stackrel{<}{\sim}
0.05$~\hmpc\ as explained in Section 2.

By construction, on large scales the HDB spectrum is identical to the
scale-invariant $\Lambda$CDM spectrum, i.e., $S(k)=1$.  Deviations
start at $k \simeq 0.03$~\hmpc: all initial spectra have a depression
around $k=0.035$~\hmpc, a bump at $k=0.05$~\hmpc, and an approximate
power law dependence on smaller scales.  The initial spectrum given in
(\ref{init}) depends only slightly on the $\sigma_8$ parameter.  For
the matter density $\Omega_m=0.3$, deviations of our accepted spectra
from the scale-invariant spectrum are moderate and localized near
$k=0.05$~\hmpc.  We call these spectra ``locally
broken-scale-invariant''.  In contrast, for the matter density
$\Omega_m=0.5$ and $\Omega_m=1.0$, deviations on small scales become
large. Even if we do not include a small fraction of hot dark matter,
the function $S(k)$ can be brought closer to unity on the scale
interval $0.06 \le k \le 0.4$~\hmpc\ using a tilted spectrum.  Tilted
spectra have been often used to explain the observed power spectrum of
galaxies. This helps to bring $S(k)\approx 1$ over the whole range
only for $\Omega_m=0.3$.  For higher values of the matter density, the
tilt needed becomes too large and is excluded using the COBE data,
that suggest $n = 1 \pm 0.1$.
 
To summarize, the CMB data in combination with the matter spectrum and
cluster abundance favor the density parameter $\Omega_m\simeq 0.3$.
For this density value, the initial matter power spectrum is
approximately scale invariant.  Figure~2 indicates that for
$\Omega_m\simeq 0.3$ the CMB data prefers $\sigma_8=0.65$ for both the
HDL and HDB spectra, while the value of $\Omega_m\simeq 0.5$ is
preferred by the data when $\sigma_8=0.75$.  However, this density
value exceeds considerably the value preferred by other data sets
(Perlmutter et al. 1998, Riess et al. 1998, see also Ostriker \&
Steinhardt 1995), and also requires large deviations from the
scale-invariant initial conditions.

\section{Conclusions}

We have shown that the amplitude of the second acoustic peak
measured by BOOMERANG-98 and MAXIMA-I is compatible both with the
standard BBN and with the matter power spectrum obtained from galaxy
and cluster catalogs, if we accept a non-scale-invariant power
spectrum.  If we assume that the age of the Universe is $14$ Gyr, and
the initial power spectrum has an index close to the
Harrison-Zeldovich value, $n=1 \pm 0.1$, then our best fit within the
standard BBN bound gives for the matter density parameter, $0.3 \le
\Omega_m \le 0.5$.  If we demand that the matter power spectrum yields
a correct cluster abundance, then the matter density is further
constraint: $\Omega_m \approx 0.3$.

The amplitude of the second peak in the radiation power spectrum
and the shape of the matter power spectrum can be both simultaneously
explained using models with the standard baryon abundance.
Since our accepted spectra have less power
than the scale-invariant model with the same cosmological parameters
(and before the new data was available) 
we had predicted a low third acoustic peak. 
This testable prediction of our models can be clearly
seen in Figs 1(e), 1(f): the third acoustic peak is not higher (and
typically lower) than the second peak.  While the original paper
was being refereed, Netterfield et al. (2001) extended the
analysis of the BOOMERANG-98 data to higher multipoles and indeed found
an amplitude of the third Doppler peak being of a similar amplitude as
the second one. However, 
Lee et al. (2001) claim to have seen a larger third peak in the
MAXIMA-I data.
In this respect, let us remark that HDL and HDB spectra coincide for
$k>0.05$~\hmpc, their CMB angular spectra are practically the same for
$l>500$ (if $\Omega_m=0.3$) while in scale invariant models with a high baryon
abundance and models with leptonic asymmetry the third peak is higher
than the second. As the data around the third peak improves,
its amplitude will be an important test on our models. 

Our second result is the shift of the first acoustic peak to a smaller
value of $l$ as compared to the scale-invariant spectrum (see Table
1).  This effect is especially noticeable in the case of the HDB
spectrum.  This brings the BOOMERANG-MAXIMA data to excellent agreement with
a spatially flat Universe, removing the need for a slightly positive
spatial curvature.  Also, the cluster mass function for the HDB type
models is well within the observational error bars.

Finally, it is important to notice that the data do not
contradict the existence of a bump and a depression of the HDB
spectrum.  The existing CMB data are not sufficient to determine the
detailed form of the matter power spectrum in the range
$0.03<k<0.06$\hmpc\ where we expect a deviation from the
scale-invariant spectrum to be most significant. 
Analysis of the data around the third peak will be useful
to determine the amplitude of the matter power spectrum at
small scales, but is not relevat for the question about the
feature. But before this analysis can be carried out, the discrepancy
between Maxima and other experiments should be resolved 
in some way, i.e., is the third peak high or low.

\vspace*{0.2cm}
\noindent{\it Acknowledgements:} We thank Mirt Gramann and Enn Saar
for discussion.  F. A.-B. acknowledges the financial support or the
Junta de Castilla y Le\'on (project SA 19/00B) and the Ministerio de
Educaci\'on y Cultura (project BFM2000-1322). J.E. was supported by
Estonian Science Foundation grant 2625. A.S. was partly supported by
the Russian Foundation for Basic Research, grant 99-02-16224, and by
the Russian Research Project ``Cosmomicrophysics''.

\end{document}